\documentclass[11pt,a4paper]{article}

\usepackage[margin=1.2in]{geometry}
\usepackage{graphicx}
\usepackage{amsmath}
\usepackage[font=small,labelfont=bf]{caption}
\usepackage{subcaption}
\usepackage{makecell}
\usepackage{cite}
\usepackage{notoccite}

\begin{document}
\date{}
\title{On the Modelling and Numerical Simulation of Non-Newtonian Blood Flow in an Aneurysm}
\maketitle

\begin{center}
\textit{Zahra Khodabakhshi Fard}\textsuperscript{1}, \textit{Azadeh Jafari}\textsuperscript{1*} \\~\\~\\

{\footnotesize \textsuperscript{1}School of Mechanical Engineering, College of Engineering, University of Tehran, Tehran, P.O. Box: 11155-4563, Iran \\
\textsuperscript{*}E-mail : azadeh.jafari@ut.ac.ir \\~\\~\\ \par}
\end{center}

\begin{center}
\textbf{Abstract} \\
\end{center}
Cardiovascular diseases, specifically cerebral aneurysms, represent a major cause of morbidity and mortality, having a significant impact on the cost and overall status of health care. In the present work, we employ a haemorheological blood model originally proposed by Owens to investigate the haemodynamics of blood flow through an aneurytic channel. This constitutive equation for whole human blood is derived using ideas drawn from temporary polymer network theory to model the aggregation and disaggregation of erythrocytes in normal human blood at different shear rates. To better understand the effect of rheological models on the haemodynamics of blood flow in cerebral aneurysms we compare our numerical results with those obtained with other rheological models such as the Carreau-Yasuda (C-Y) model. The results show that the velocity profiles for the Newtonian and the Owens models are approximately similar but differ from those of the C-Y model. In order to stabilize our numerical simulations, we propose two new stabilization techniques, the so-called N-Owens and I-Owens methods. Employing the N-Owens stabilization method enables us to capture the effect of erythrocyte aggregation in blood flow through a cerebral aneurysm at higher Weissenberg ($We$) and Reynolds ($Re$) numbers than would otherwise be possible. \\~\\
{\footnotesize \textbf{Keywords} : \textit{Haemorheological models, red blood cell aggregation, numerical instability} \par}
\newpage

\subsection*{Nomenclature}
\begin{center}
\begin{tabular}{ l l }
Symbol & Description \\
\hline\\
 $|\dot\gamma | [s^{-1}]$ & shear rate \\ 
 $\mu [Pa.s]$ & Newtonian fluid viscosity\\ 
 $\mu_s [Pa.s]$ & plasma viscosity\\
 $\mu_0 [Pa.s]$ & zero shear rate viscosity\\ 
 $\mu_\infty [Pa.s]$ & infinite shear rate viscosity\\ 
 $\mu_e [Pa.s]$ & effective (or apparent) viscosity\\ 
 $\tau$ [Pa] & stress tensor\\
 $n$ & average RBC aggregate size\\
 $v [m.s^{-1}]$ & velocity vector \\
 $\lambda_H [s]$ & relaxation time of a single RBC\\
 $\lambda [s]$ & fluid relaxation time \\

\end{tabular}
\end{center}

\section{Introduction}
An aneurysm is the extreme widening of blood vessel walls \cite{ChristopherL.TaylorZhongYuanWarrenR.SelmanRobertA.Ratcheson1995}. Aneurysms may occur in many parts of the cardiovascular system like the aorta \cite{Finol2001a}, cerebral arteries, eye microvessels, etc \cite{Ballantyne1944}. Usually aneurysms can be detected by Magnetic Resonance Imaging (MRI) or Computed Tomography Angiography (CTA), but in a cerebral aneurysm the procedure is somewhat more difficult. In most cases, MRI can't detect aneurysms which are smaller than 5 mm in diameter and usually alternative methods such as CT-angiography or Magnetic Resonance Angiography (MRA) may be used \cite{Wardlaw2000a}. Treatment of such small aneurysms needs complex surgery with a high morbidity rate \cite{Stringfellow1987}. Increasing demand from the medical community for scientifically exact quantitative investigations of vascular diseases, has recently given a major impetus to the development of mathematical models and numerical tools for computer simulations of the human cardiovascular system, specifically on cerebral aneurysms, in both healthy and pathological states \cite{Ortega1998}.

From a macroscopic viewpoint, human blood can be regarded as a homogeneous fluid. However, at the microscopic level, blood is a suspension in which blood cells such as red blood cells (RBCs), white blood cells (WBCs) and platelets are suspended in an aqueous polymer solution. Plasma, contains electrolytes and organic molecules such as metabolites, hormones, enzymes, antibodies and other proteins \cite{DINTENFASS1963,Thurston1979}. For specialists in fluid mechanics, detailed information of complex behaviour in the cardiovascular system is of the utmost importance in any attempt at establishing the equations that govern the flow of blood in various parts of the circulatory system in different states. Depending on the shear rate of a given flow, the nature of blood can be categorized as being either Newtonian or non-Newtonian. The assumption of Newtonian behaviour is acceptable in high shear rate flows, such as flow through larger arteries. It is not, however, valid when the shear rate is low; typically less than $100 s^{-1}$. It should be emphasized that blood flow is Newtonian in most parts of the arterial system \cite{Merrill} and, if non-Newtonian effects are to be observed, attention should therefore be given to other flow regimes and clinical situations. These include, for normal blood, sluggish flow in the venous system and parts of the arterial vasculature where the geometry has been altered and RBC aggregates become more stable, such as downstream of a stenosis \cite{Chen2006} or inside a saccular aneurysm \cite{Ortega1998}. In addition, several pathologies are accompanied by significant changes in the mechanical properties of blood resulting in alterations in blood viscosity and viscoelastic properties, as reported in a number of review articles \cite{Valencia2006}. The non-Newtonian behaviour of blood is mainly caused by three phenomena: the tendency of erythrocytes to form three-dimensional micro-structures (rouleaux) at low shear rates, their deformability and the tendency of rouleaux to break up and align with the flow field at higher shear rates \cite{Cokelet1963}. Haemorheological studies have documented three types of non-Newtonian blood properties: thixotropy, viscoelasticity and shear thinning \cite{Thurston1979}, which are influenced by the level of aggregation of red blood cells, plasma viscosity and haematocrit \cite{Campo-Deano2013}.

For the simulation of blood flows involving the possible growth and risk of rupture in aneurysms, there are some important parameters to determine in the description, for example, of the geometry \cite{Nader-Sepahi2004,Ujiie1999}, haemorheological blood model \cite{Glagov1988,PEDERSEN19931237} and fluid-solid interaction \cite{Raghavan2005}. Studies demonstrate that low wall shear stresses  may enfeeble vessel wall cells and escalate the risk of an aneurysm rupturing \cite{Chappell1998,Davies1986, Helmlinger1991}. Also, a wall shear stress exhibiting a high gradient will damage vessel wall structure \cite{Ku1985,Nagel1999,Zhao2002}. Therefore, a large number of numerical and experimental studies have been performed to investigate wall shear stress effects on aneurysm rupture risk \cite{Finol2001,Morris2004,Salsac2006,Yu1999}. Some studies investigate the effect of sharp bends and bifurcation on the  shear stress distribution \cite{BuchananJr.1999,Friedman1984,Liepsch1984,Shipkowitz2000}. Buchanan et al. \cite{BuchananJr.1999} and Moore et al. \cite{Moore2006} simulated blood flows in cerebral arteries networks in three dimensions using MRI data. A realistic haemorheological model of blood is one of the most important ingredients in the simulation of a cerebral aneurysm. The C-Y blood model has been used to simulate blood flow through a cerebral aneurysm \cite{Bernsdorf2009,OBernabeu2013}. It was shown that wall shear stress (WSS) in the C-Y model is lower than that predicted on the basis of the Navier-Stokes equations and the Newtonian model may overestimate WSS in low Reynolds number flows \cite{OBernabeu2013}. Morales et al. \cite{Morales2013} investigated the interaction between coils and blood flow in a coiled aneurysm. They used the Casson and Newtonian models. The Newtonian model was found to overestimate the velocity magnitude in the coiled aneurysm without changing the overall flow patterns.

The aim of the present study is to investigate the effects of haemorheological models and parameters such as blood viscosity, haematocrit, heart rate, blood pressure, etc. on flow patterns, velocity magnitude and WSS in the cerebral aneurysm region. 
\section{Governing equations}
In this study we compare the predictions in flow through an aneurysm using Newtonian, C-Y and the homogeneous Owens \cite{Owens2006} models. The Owens model is a relatively simple haemorheological model able to predict aggregation and disaggregation of red blood cells correctly.  The mathematical description of incompressible non-Newtonian blood flows involves equations governing the conservation of mass and linear momentum as well as a suitable constitutive equation. 

The equations of conservation of mass and linear momentum are given, respectively, by
\begin{equation}
\nabla\cdot v=0
\label{eq00}
\end{equation} 
\begin{equation}
\rho \frac{Dv}{Dt} = - \nabla p +  \nabla\cdot\tau
\label{eq01}
\end{equation}
where $v$, $p$ and $\tau$ denote the velocity, pressure and stress fields, respectively. The stress tensor $\tau$
\begin{equation}
\tau =   \tau _{s} +   \tau _{p}
\label{eq02}
\end{equation}
consists of the sum of the solvent stress ($\tau _{s}$, due to the plasma),  and the polymeric stress ($\tau _{p}$) and in the Owens model \cite{Owens2006} the latter satisfies the rate equation 
\begin{equation}
\tau_p +  \lambda  \left( \frac{ \partial  \tau_p}{ \partial t} +  v  .  \nabla  \tau_p -  \nabla  v  .  \tau_p -  \tau_p  .   \nabla  v ^{T}  \right) =  \frac{  \mu _{ \infty } }{  \lambda _{H} }  \lambda  \dot{ \gamma}
\label{eq03}
\end{equation}
$\dot\gamma$, the shear rate tensor, is calculated from
\begin{equation}
\dot{ \gamma} = \nabla v +   \nabla v^{T}
\end{equation}
For more detailed information about the Owens model the reader is referred to Appendix A and to the original paper \cite{Owens2006}.

The numerical solution of the equations governing the C-Y and Newtonian model flows is relatively straightforward. The polymeric stress $\tau_p$  equals zero and the Newtonian stress is determined from 
\begin{equation}
\tau_N=  \mu  \dot{ \gamma}
\label{eq04}
\end{equation}	
For Newtonian blood flows, the viscosity $\mu$ equals the high shear-rate whole blood viscosity which is $0.0035 Pa.s$  and for the C-Y model the effective viscosity depends upon the shear rate in the following way: 
\begin{equation}
\mu = \mu_e =  \mu_\infty +  \frac{\mu_0 - \mu_\infty }{\left(1+  \left( \lambda  |  \dot{ \gamma }  | \right) ^{B} \right) ^A} 
\label{eq05}
\end{equation}
The various parameters in the C-Y model may be taken to equal \cite{Nichols1998, Armstrong1988}: 
\begin{equation}
\mu_\infty = 0.0035 Pa.s,\mu_0 = 0.16 Pa.s, \lambda = 8.2 s, B=1.23, A=0.64 
\label{eq06}
\end{equation} 
At sufficiently high shear rate flow the effective viscosity in the C-Y model tends to the high shear rate blood viscosity. %
\section{Geometry and Spatial Discretization} 
\begin{figure}
\centering
\includegraphics[scale=0.75]{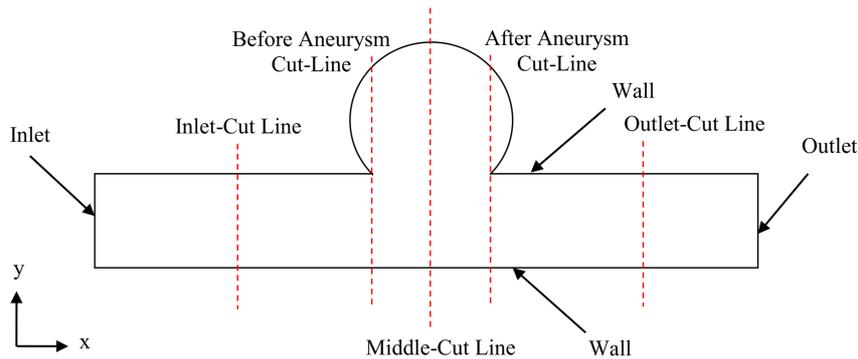}
\caption{Schematic of geometry of vessel with aneurysm}
\label{Fig1}
\end{figure}

\begin{figure}
\centering
\includegraphics[scale=0.9]{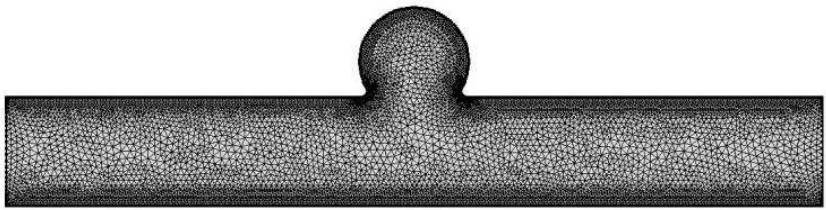}
\caption{Mesh having 15000 triangular elements}
\label{Fig2}
\end{figure}

In this paper we will present numerical results for blood flow through a two dimensional aneurytic channel under
both steady and pulsatile flow conditions. The geometry is shown in Fig \ref{Fig1}. The numerical method employed for all computations was the finite element method.  To enhance the stability of our computations, we employed streamline upwind stabilization.  The linearized system of equations was solved at each time step using a fully coupled direct method. Spatial discretization of both the velocity and pressure fields was based on $P_1$ triangular elements. Note that for non-Newtonian simulations using the Owens model second-order $P_2$ triangular elements were used to calculate the viscoelastic stress. All computations were performed in rigid-walled channels. A mesh having 15000 triangular elements is shown in Fig. \ref{Fig2}. Refinement of the mesh near the vessel wall and the aneurysm sac is based on an advancing front Delaunay triangulation algorithm \cite{Mavriplis1995}. Both corners of the aneurysm sac have been rounded off in order to reduce the effect of what would otherwise be singularities at the corners.
\section{Boundary Conditions} 
For the simulation of blood flow in the aneurytic channel with the Newtonian and C-Y models, we imposed a fully developed velocity profile of the Newtonian fluid at the inlet boundary and a constant pressure at the outlet boundary. To simplify the implementation of the problem, we also used a Newtonian fully developed velocity profile for the C-Y model at the inlet boundary. No slip boundary conditions were imposed on the vessel wall. The stability of the numerical computations using the Owens model is highly dependent on the choice of boundary conditions.  The fully developed velocity and stress profiles were imposed at the inlet with the assumption that the shear rates were high enough to allow us to ignore the variation of polymeric viscosity with shear rate at the inflow boundary. The analytical results are shown in Appendix B. No slip and zero normal stress boundary conditions were imposed on the vessel walls. We assumed that the outlet boundary was far enough downstream of the aneurysm that the flow became fully developed once again and that the axial gradients of velocity and stresses could be taken to be zero at outflow. The initial conditions were based on the inflow boundary values. The initial value for the aggregation size for the RBCs was set equal to $1$ (i.e. the initial state consisted of disaggregated RBCs).

\begin{figure}
\centering
\includegraphics[scale=1]{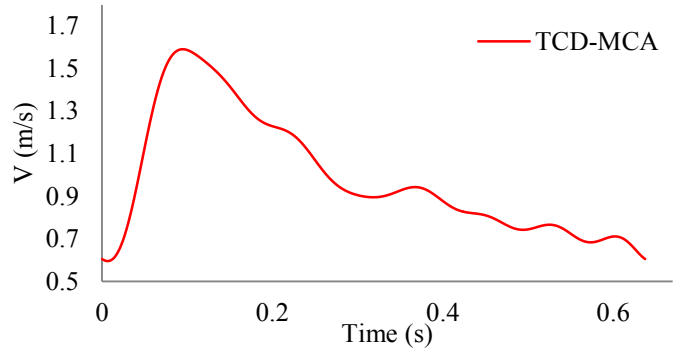}
\caption{The maximum value of the inlet velocity at different times, measured at the MCA position \cite{Wagshul2011}}
\label{Fig3}
\end{figure}

In our pulsatile blood flow, the maximum velocity of the parabolic velocity profile at inflow was chosen to equal the experimental values obtained by Transcranial Doppler (TCD) at the middle cerebral artery (MCA) position \cite{Wagshul2011} as shown in Fig. \ref{Fig3}.
%
\section{Validation}
\subsection{Comparison with the results of Owens}
\begin{figure}
\centering
\includegraphics[scale=0.7]{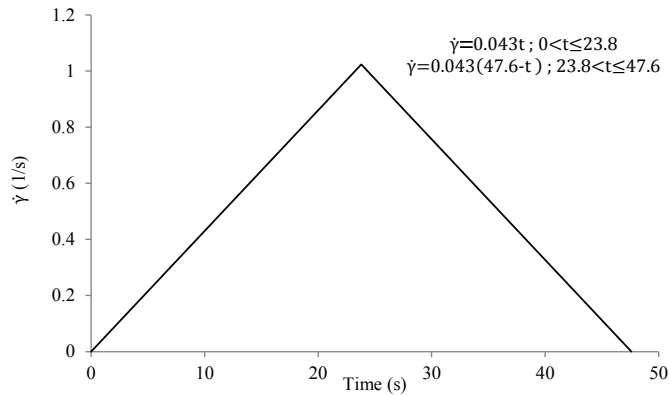}
\caption{Triangular step shear-rate experiment}
\label{Fig4}
\end{figure}

\begin{figure}
\centering
\includegraphics[scale=0.75]{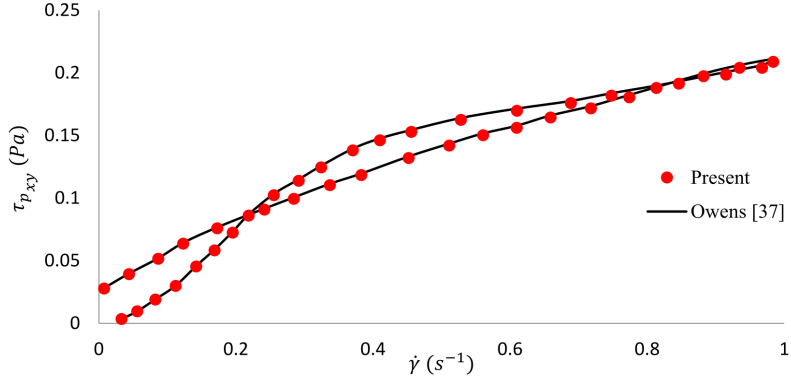}
\caption{Variation of shear stress versus shear rate in hysteresis experiment.}
\label{Fig5}
\end{figure}

In order to show the validity of our code, we consider a triangular step shear-rate experiment in a narrow gap coaxial cylinder microviscometer originally performed by Bureau et al. \cite{Bureau1980}. The experiment, Fig. \ref{Fig4}, was done in such a way that the shear rate increases linearly to its maximum value over a time interval $[0, t_0]$  followed by a linear return to zero at time $2t_0$. The boundary conditions are the same as those in Owens's article \cite{Owens2006}. As may be seen in Fig. \ref{Fig5}, the results of our numerical simulation are in good agreement with those of Owens \cite{Owens2006}. 
\subsection{Convergence with mesh refinement: steady non-Newtonian flow through an aneurytic channel}
\begin{figure}
\centering
\begin{subfigure}{.5\textwidth}
  \centering
  \caption{}
  \includegraphics[width=.8\linewidth]{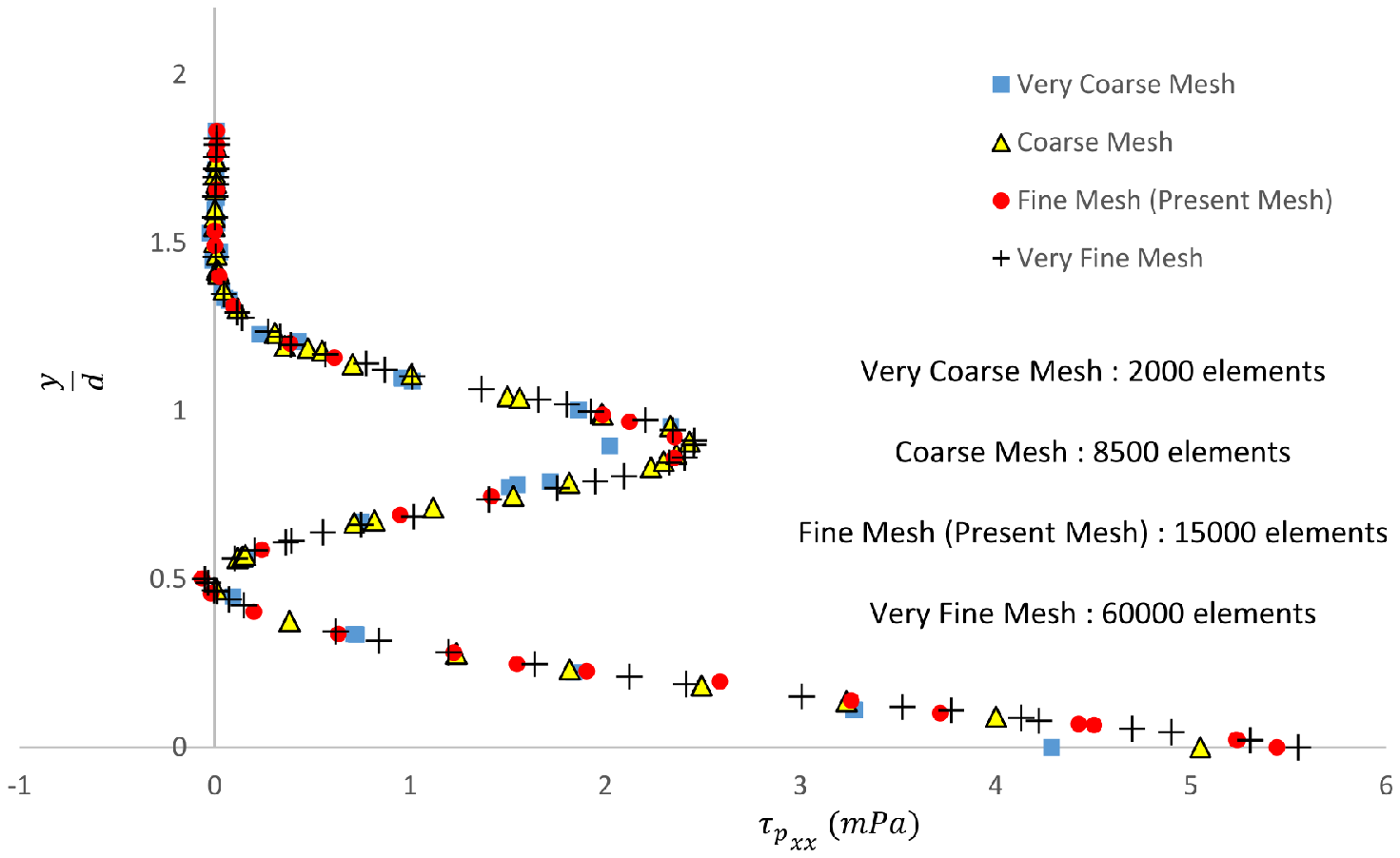}
  \label{Fig12_a}
\end{subfigure}%
\begin{subfigure}{.5\textwidth}
  \centering
  \caption{}
  \includegraphics[width=.8\linewidth]{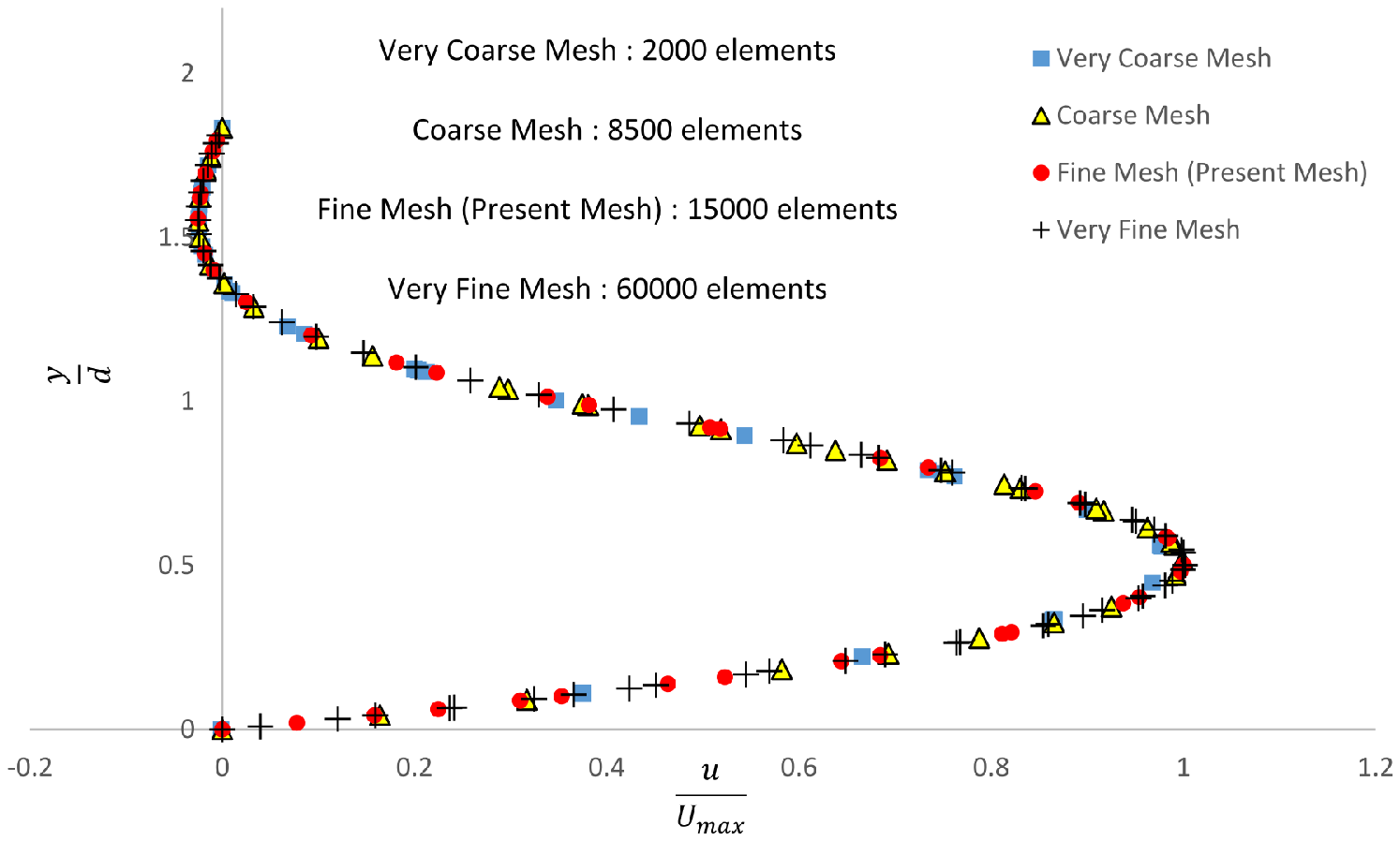}
  \label{Fig12_b}
  \end{subfigure}
\caption{Comparison of the (a) first normal polymeric stress ${\tau_p}_{xx}$ and (b) velocity profile along the middle-cut line (shown in Fig. \ref{Fig1}) with different meshes}
\label{Fig2prime}
\vspace{-1mm}
\end{figure}

The results, in the case of steady non-Newtonian flow in an aneurytic channel, of a study of convergence with mesh refinement are depicted in Fig. \ref{Fig2prime}. Here we compare the normal polymeric shear stress ${\tau_p}_{xy}$ and velocity profile along the middle-cut line using four different meshes. Very coarse, coarse, fine and very fine meshes, having 2000, 8500, 15000 and 60000 triangular elements, respectively, were used. From this study it would seem that the mesh having 15000 elements (see Fig. \ref{Fig2}) is adequate for capturing the haemodynamics of blood flow in an aneurytic channel. Consequently, this mesh will be the one used for all the computations of Sections \ref{sec:7} and \ref{sec:8}. 

 \captionsetup[subfigure]{position=top,singlelinecheck=off,justification=raggedright}
\section{Results}\label{sec:7}
\begin{figure}
\centering
\includegraphics[scale=0.7]{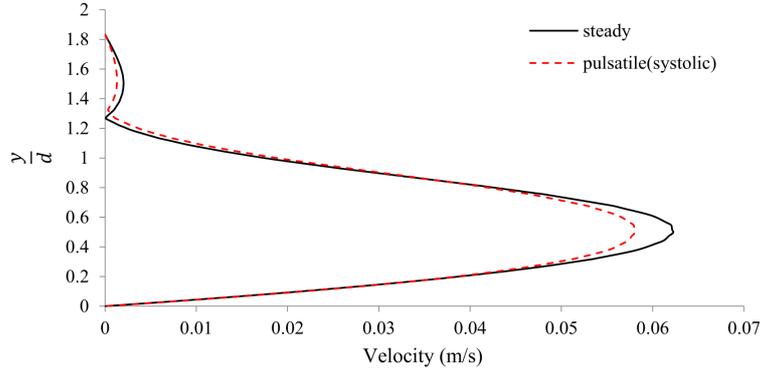}
\caption{Velocity profile along the middle cut line, which is shown in Fig. \ref{Fig1}, for both steady and pulsatile flows (systole)}.
\label{Fig6}
\end{figure}

\begin{figure}
\centering
\includegraphics[scale=0.7]{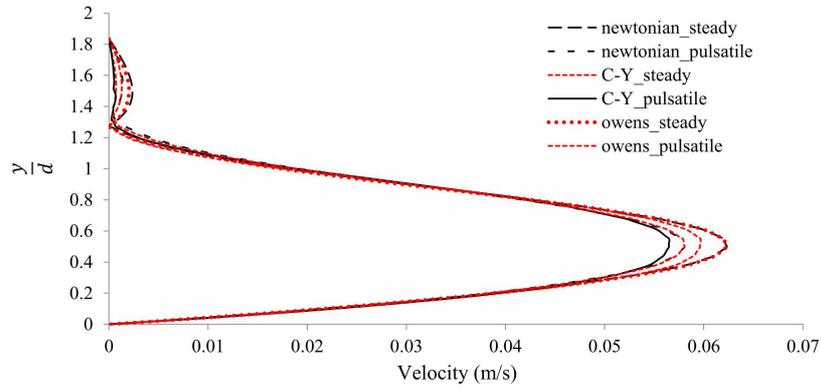}
\caption{Comparison of velocity profiles with the Owens, C-Y and Newtonian models in steady and pulsatile flows (systolic state) in the middle cut line which is shown in Fig. \ref{Fig1}}
\label{Fig7}
\end{figure}

\begin{figure}
\centering
\includegraphics[scale=0.7]{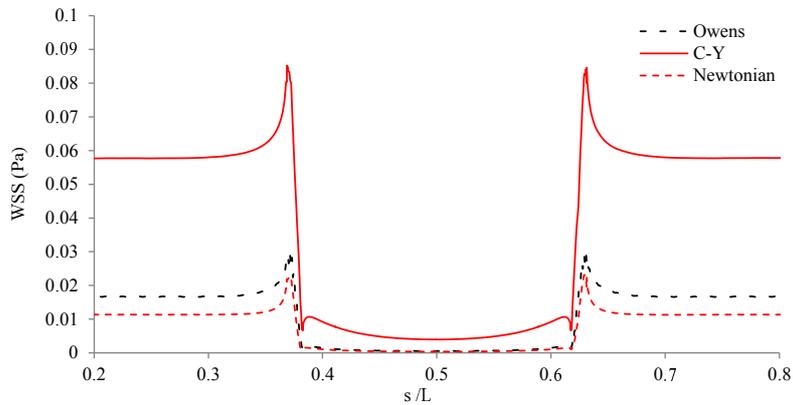}
\caption{WSS on the upper wall of the channel computed with the Owens, C-Y and Newtonian models in steady state at $Re=10$. $L$ denotes the arc length of the upper wall. The exact interval of $s/L$ values that comprise the aneurysm from the left junction of the upper wall and the neck to the right junction with the neck is $0.35$ to $0.65$.}
\label{Fig9}
\end{figure}

\begin{figure}
\centering
\includegraphics[scale=0.7]{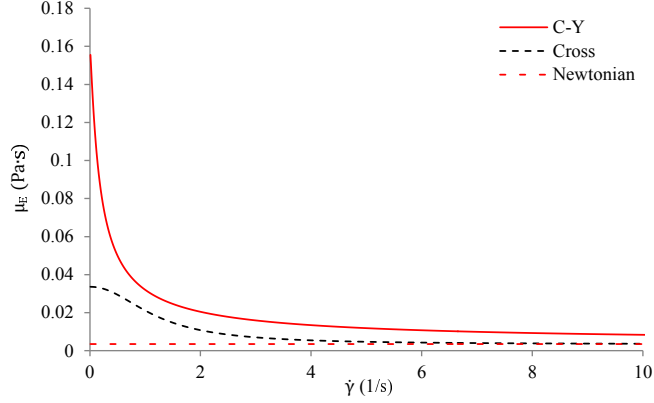}
\caption{Effective viscosity variations with shear rate at C-Y, Cross and Newtonian models}
\label{Fig8}
\end{figure}

\begin{figure}
\centering
\includegraphics[scale=0.6]{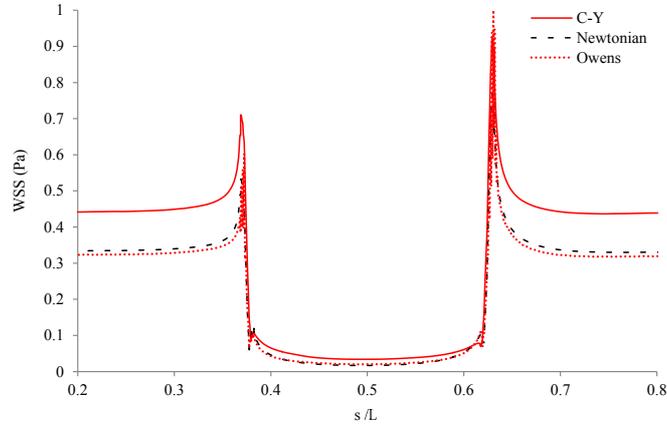}
\caption{Wall shear stress along the upper wall of the aneurysm with the Owens, C-Y and Newtonian models in diastole at $Re=30$. The exact interval of $s/L$ values that comprise the aneurysm from the left junction with the neck to the right junction with the neck is $0.35$ to $0.65$.}
\label{Fig10}
\end{figure}

\begin{figure}
\centering
\includegraphics[scale=0.6]{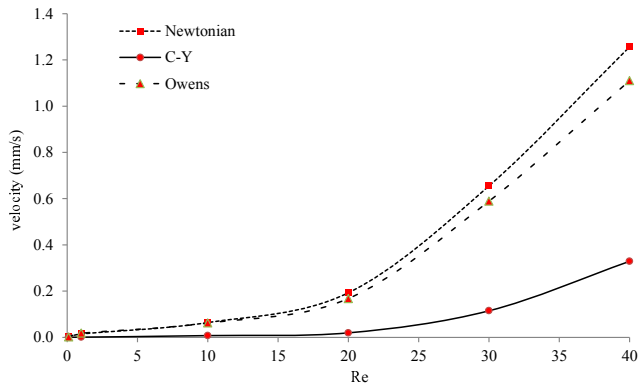}
\caption{The variation of velocity with respect to Reynolds number in the center of aneurysm sac with Owens, C-Y and Newtonian models}
\label{Fig11}
\end{figure}

\begin{figure}
\begin{subfigure}{.5\textwidth}
  \centering
  \caption{}
  \includegraphics[width=.8\linewidth]{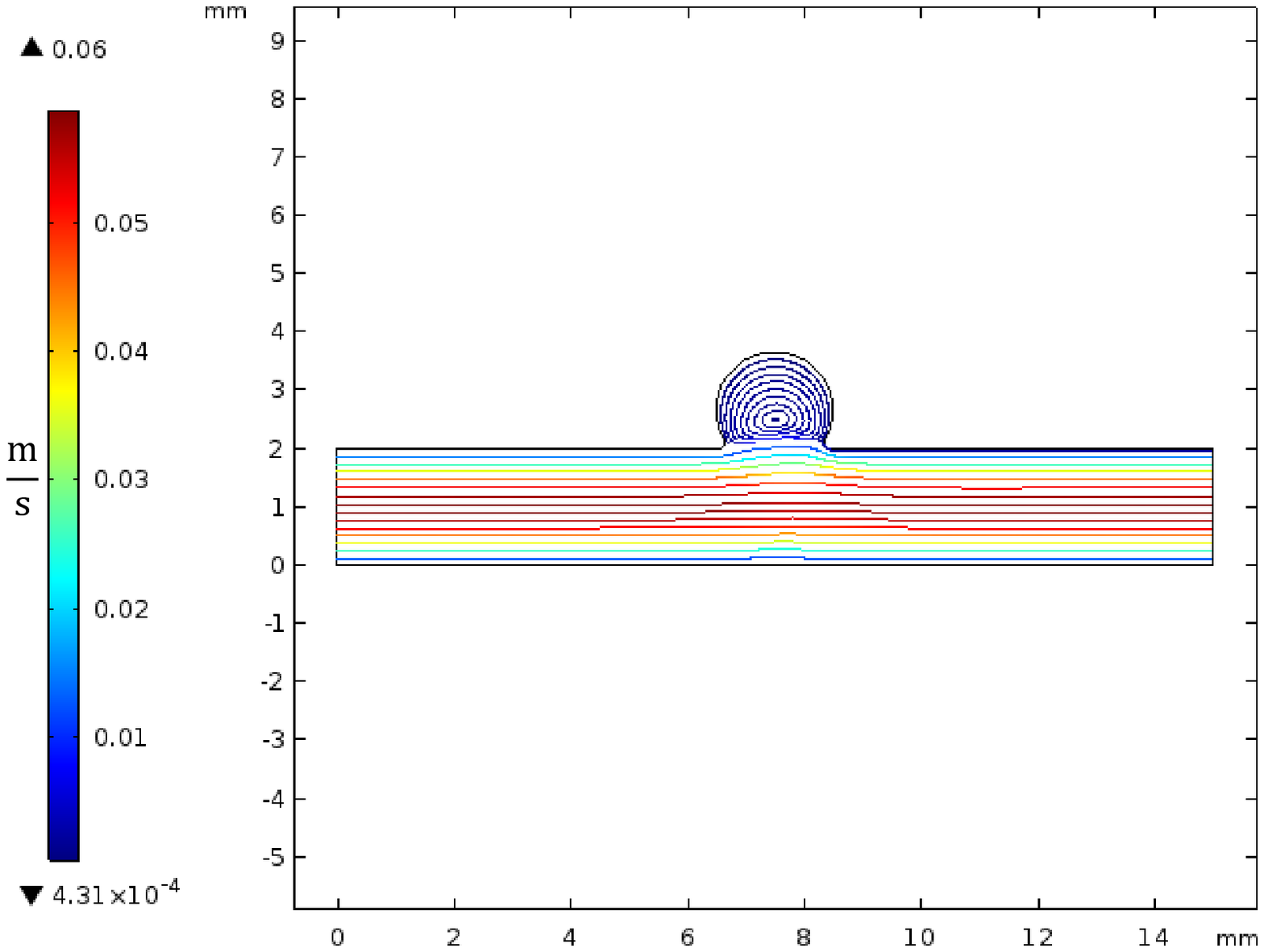}
  \label{Fig12_a}
\end{subfigure}%
\quad
\begin{subfigure}{.5\textwidth}
  \centering
  \caption{}
  \includegraphics[width=.8\linewidth]{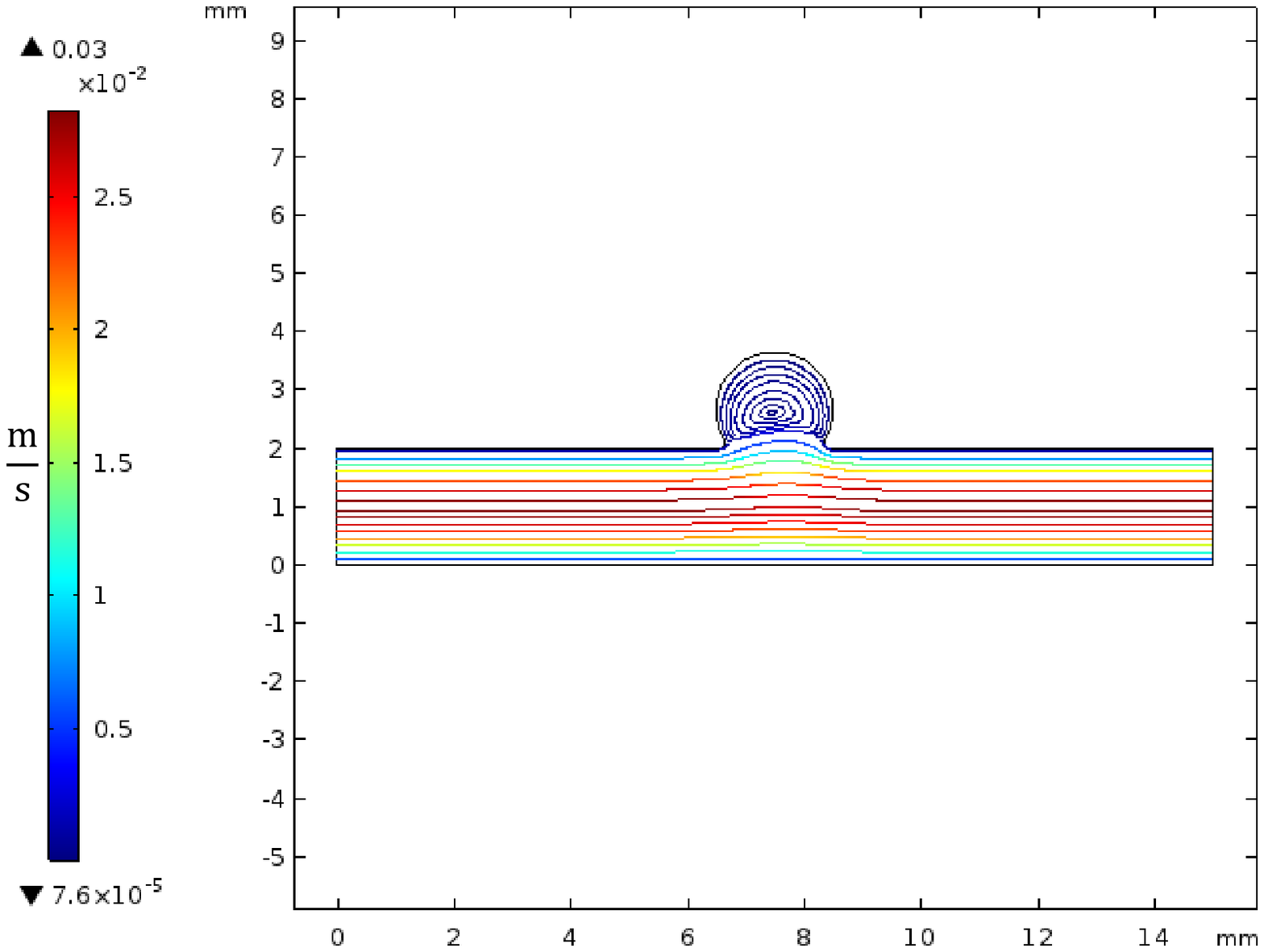}
  \label{Fig12_b}
\end{subfigure}
\quad
\begin{subfigure}{.5\textwidth}
  \centering
  \caption{}
  \includegraphics[width=.8\linewidth]{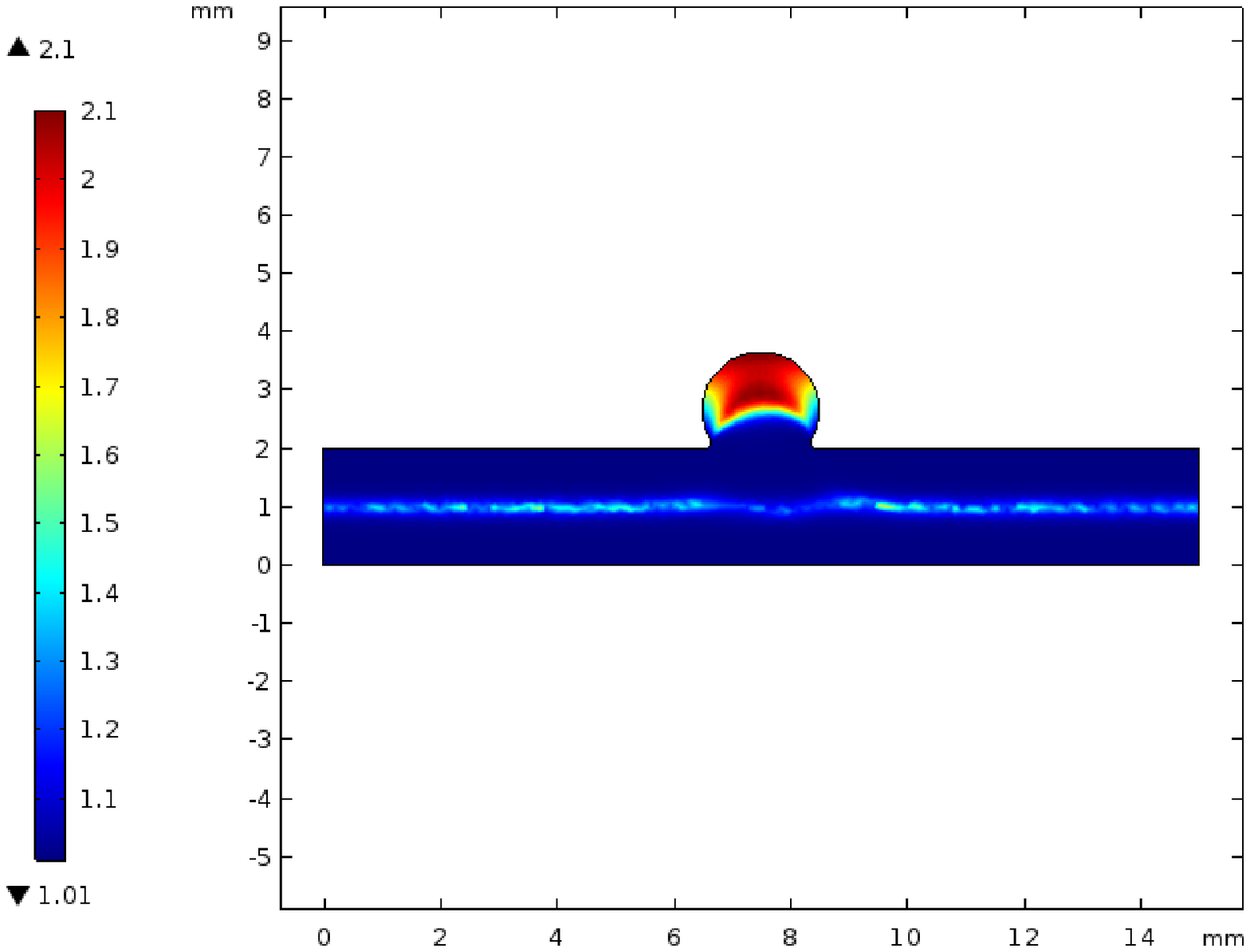}
  \label{Fig12_c}
\end{subfigure}
\quad
\begin{subfigure}{.5\textwidth}
  \centering
  \caption{}
  \includegraphics[width=.8\linewidth]{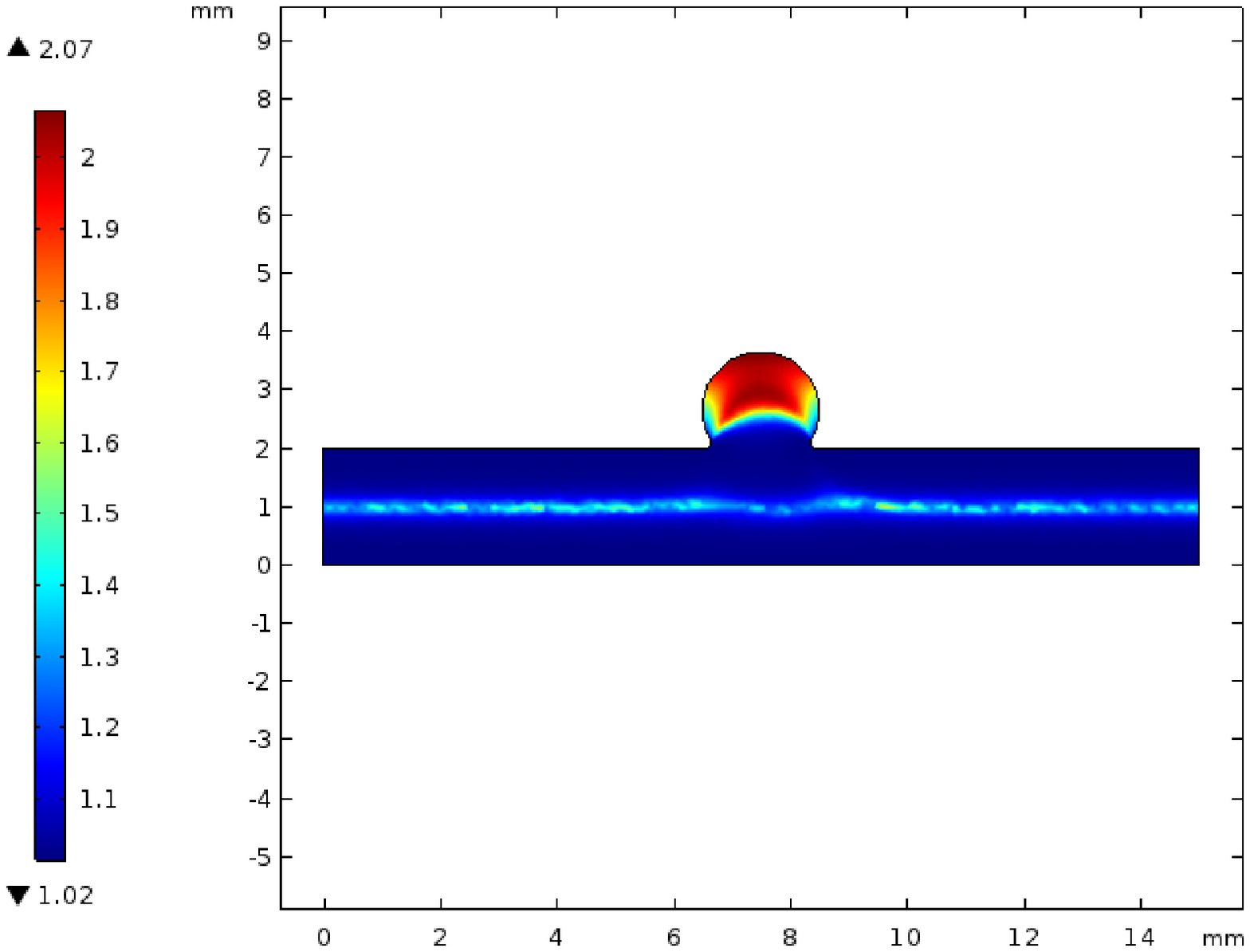}
  \label{Fig12_d}
\end{subfigure}
\quad
\begin{subfigure}{.5\textwidth}
  \centering
  \caption{}
  \includegraphics[width=.8\linewidth]{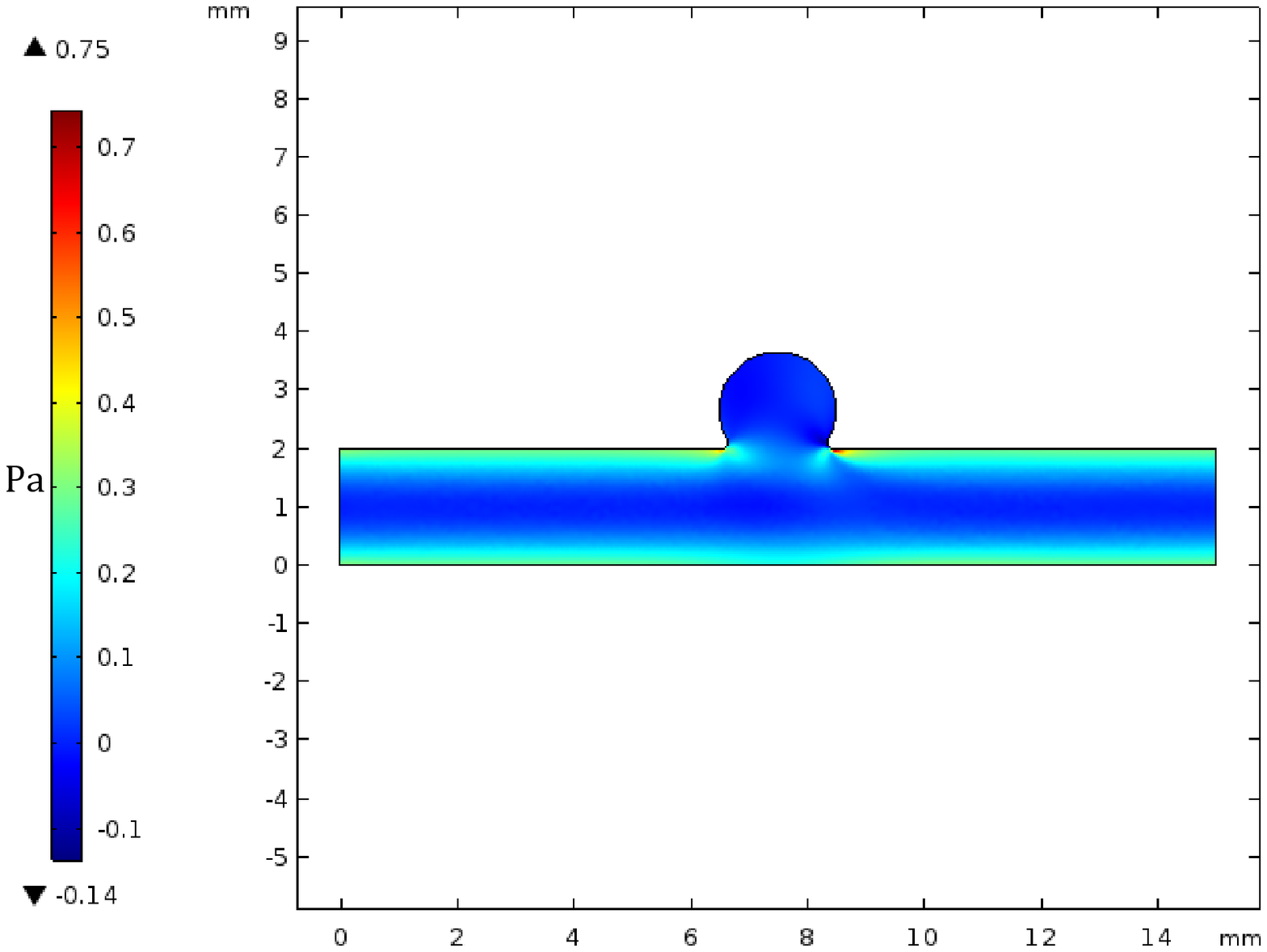}
  \label{Fig12_e}
\end{subfigure}
\quad
\begin{subfigure}{.5\textwidth}
  \centering
  \caption{}
  \includegraphics[width=.8\linewidth]{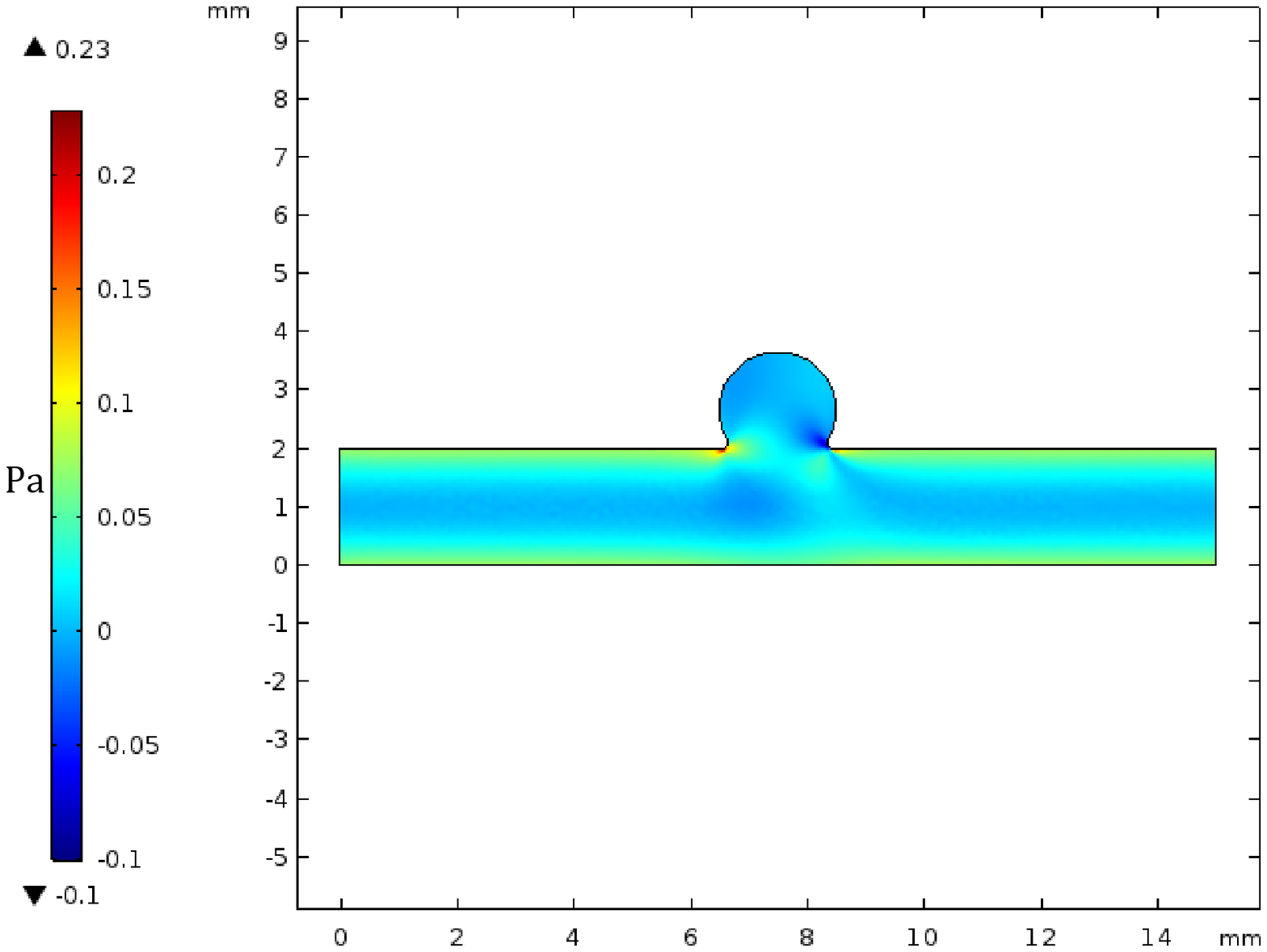}
  \label{Fig12_f}
\end{subfigure}
\quad
\begin{subfigure}{.5\textwidth}
  \centering
  \caption{}
  \includegraphics[width=.8\linewidth]{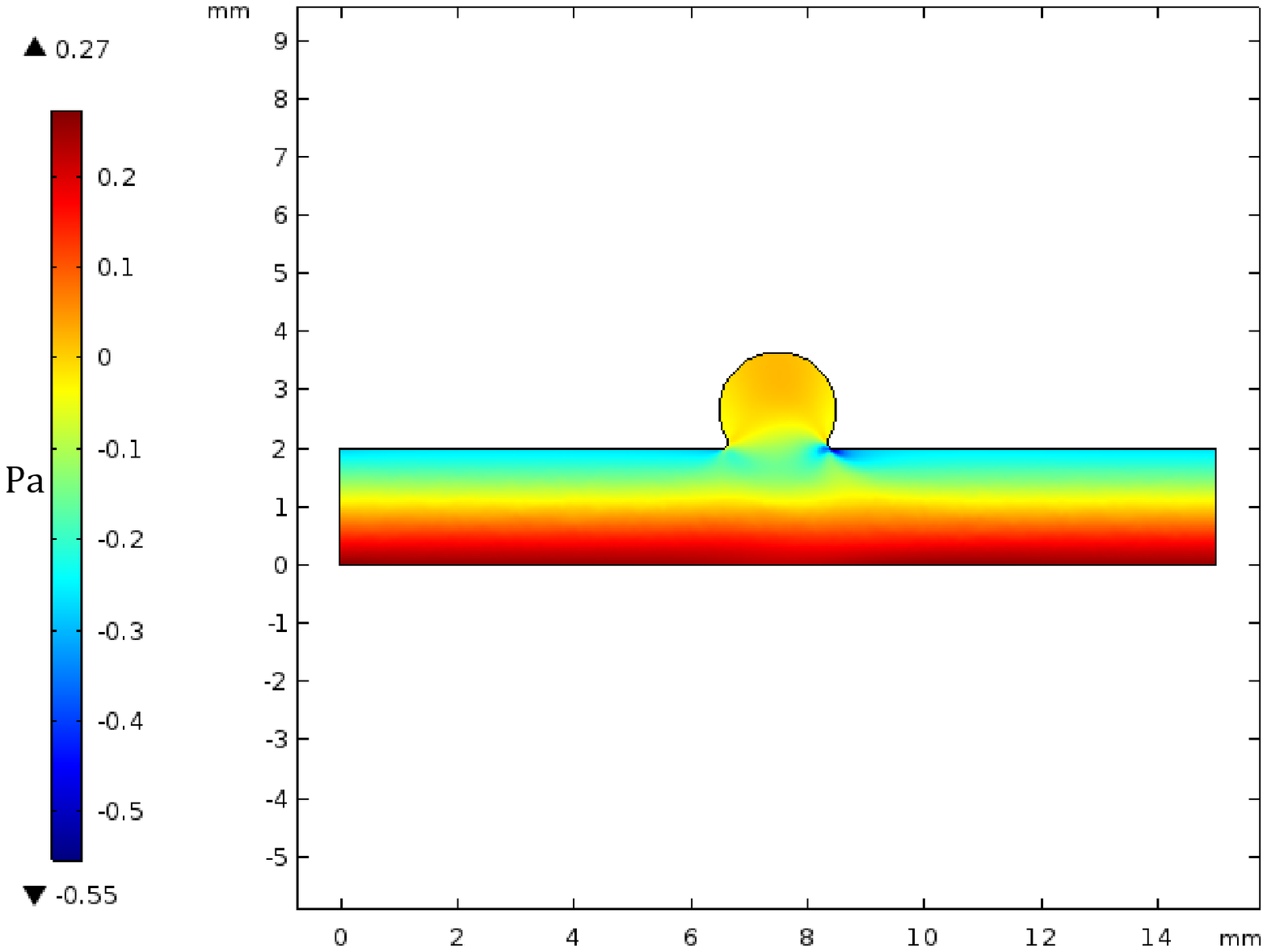}
  \label{Fig12_g}
\end{subfigure}
\quad
\begin{subfigure}{.5\textwidth}
  \centering
  \caption{}
  \includegraphics[width=.8\linewidth]{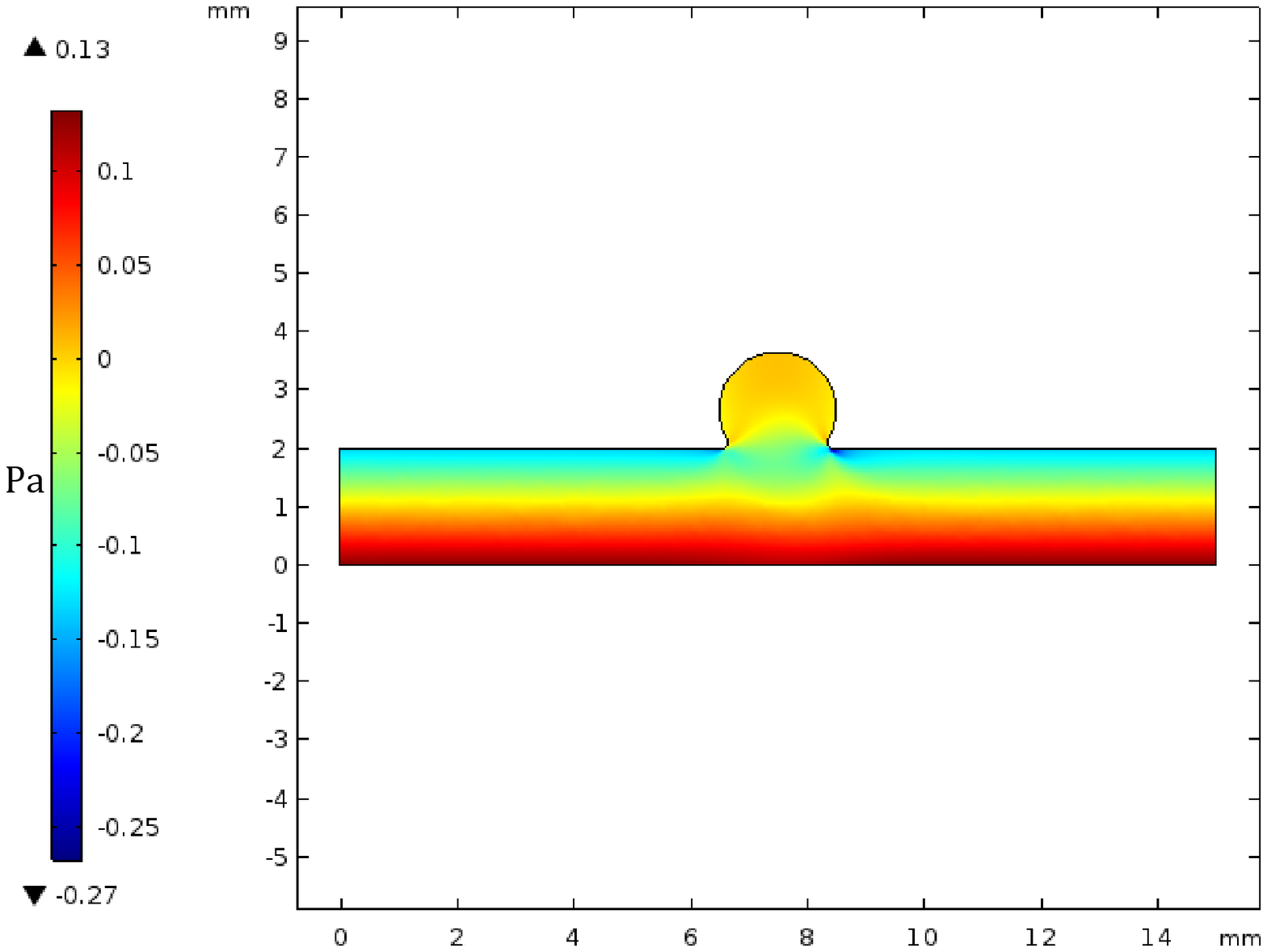}
  \label{Fig12_h}
\end{subfigure}
\caption{Streamlines (a, b), average aggregation size(c, d), ${\tau_p}_{xx}$ (e, f) and ${\tau_p}_{xy}$  (g, h) in the aneurytic channel with the Owens model in systolic (left-hand side) and diastolic (right-hand side) states}
\label{Fig12}
\end{figure}

Fig. \ref{Fig6} shows the velocity profile in the middle of the aneurysm with the Owens model for both steady and pulsatile flows at systolic pressure and at the same Reynolds number. The results of the pulsatile flow are presented after 5 cycles have elapsed since the initial conditions, by which time something approaching a steady state is achieved in a periodic flow at systolic pressure. However, due to inertial effects in pulsatile flow, the velocity profiles in both steady and pulsatile flows are not exactly the same even at the same Reynolds number \cite{Womersley1955}. 
Fig. \ref{Fig7} shows the velocity profiles of the Owens, C-Y and Newtonian models for both steady and pulsatile flows at systolic pressure. As may be observed, the velocity profiles of the Newtonian model are in good agreement with those of the Owens model for both steady and pulsatile flows, but the C-Y model shows smaller maximum values in the centre line of the channel.\\ 
Fig. \ref{Fig9} shows $WSS$ on the upper wall of the vessel at $Re=10$ and is calculated from

\begin{equation}
\begin{split}
WSS ^{2} = & \left( \left(  {\tau_p}_{xx}+2  \mu _{s} \frac{ \partial u}{ \partial x} \right) \cdot  n_{x}+  \left({\tau_p}_{xy}+  \mu _{s} \left( \frac{ \partial u}{ \partial y}+ \frac{ \partial v}{ \partial x}  \right) \right) \cdot  n_{y}\right) ^{2} \\
& + \left( \left({\tau_p}_{yy}+  2\mu _{s} \frac{ \partial v}{ \partial y} \right) \cdot n_{y}+  \left({\tau_p} _{xy}+  \mu _{s} \left( \frac{ \partial u}{ \partial y}+ \frac{ \partial v}{ \partial x}  \right) \right) \cdot n_{x}\right) ^{2}  ,
\label{eq07}
\end{split}
\end{equation}
where $n_x$ and $n_y$ denote the $x-$ and $y-$ components of the outward pointing unit normal vector on the vessel wall. 
WSS has its maximum value in the neck of the aneurysm. Although the velocity profile is approximately the same in the Owens and Newtonian models, the WSS is different in these two models. Note that the Owens model uses the Cross viscosity model to define the effective viscosity
\begin{equation}
\mu _{e}= \mu _{0} \left( \frac{1+ \phi|\dot{ \gamma }|^{m}}{1+\beta |\dot{ \gamma }|^{m}} \right) +  \mu _{s},
\label{eq08}
\end{equation}
where the ratio $\phi/\beta$ equals $\mu_\infty/\mu_0$, the ratio of the polymeric contribution to the infinite and zero shear-rate viscosities. 
Fig. \ref{Fig8} shows that the effective viscosities of the three models have significant differences at low ($|\dot\gamma |<2 s^{-1}$) shear rates. The highest value of the effective viscosity belongs to the C-Y model, so that the WSS of the C-Y model should be greater than that of the Owens and Newtonian models. 
From Fig. \ref{Fig10} it may be seen that increasing the Reynolds number to $30$ results in a significant decrease in the difference between the Owens and Newtonian model predictions of the WSS. This is because the effective viscosity of the Owens model tends to the Newtonian viscosity at high shear rates. 
Fig. \ref{Fig11} shows the velocity magnitude in the centre of the aneurysm sac with respect to Reynolds numbers for the three models. As we can see the slope of the graphs and the value of the velocity magnitude for both the Newtonian and Owens models do not change significantly with increasing Reynolds number. So the effect of the Owens model on the velocity field would seem to be negligible compared to the Newtonian value and this observation leads us to one possibility (see later) for stabilizing computations with the Owens model in higher Reynolds/Weissenberg number flows.
Fig. \ref{Fig12} shows the instantaneous streamlines, the contours of average aggregate size, $n$, the first normal polymeric stress component ${\tau_p}_{xx}$, and the polymeric shear stress ${\tau_p}_{xy}$, for both systolic (left-hand side) and diastolic (right-hand side) states in the aneurytic channel. 
Figs.\ref{Fig12}(a,b) show that as the velocity changes from the systolic to the diastolic state, the centre of the vortex moves towards the upper left corner of the aneurysm. The shifting of the vortex centre causes the WSS to increase at the upper right corner of the aneurysm, as depicted in Fig. \ref{Fig10}. The RBC aggregation zone in both systolic and diastolic states is illustrated in Figs.\ref{Fig12}(c,d). $n$, the average aggregate size, increases in the lower shear-rate region of the centreline of the blood vessel and the aneurysm sac. Consequently, the results show the ability of the Owens model to capture RBC aggregation. The distribution of the first normal polymeric stress, ${\tau_p}_{xx}$ and the shear stress ${\tau_p}_{xy}$, is shown in Figs.\ref{Fig12}(e,f) and Figs.\ref{Fig12}(g,h), respectively. It is obvious that the polymeric shear stress in the aneurysm sac is strongly influenced by the drastic variations in the shear-rate $|\dot\gamma |$ and the average aggregate size $n$.
\captionsetup[subfigure]{position=top,singlelinecheck=off,justification=raggedright}
Table \ref{Table1} shows a comparison of the predictions of the mean velocity in the MCA using our 2D homogeneous Owens model with the physiological data from Pott et al. \cite{Pott1997}, Valdueza et al. \cite{Valdueza1997}  and Lindegaard et al. \cite{Lindegaard1987}. In all cases the average radius of the parent vessel was around 1mm.  Particularly good agreement between the numerical predictions and the physiological data of Lindegaard et al. \cite{Lindegaard1987} may be observed.

\begin{table}[]
\centering
\caption{comparison of 2D aneurysm model with physiological data in the MCA} \label{Table1}
\begin{tabular}{|c|c|c|cc|}
\hline
\begin{tabular}[c]{@{}l@{}}Model \end{tabular}              & \begin{tabular}[c]{@{}l@{}}Mean Radius\\   (mm)\end{tabular} & \begin{tabular}[c]{@{}l@{}}Mean Velocity\\   (m/s)\end{tabular} & \begin{tabular}[c]{@{}l@{}}Deviation of\\   2D model\end{tabular} &  \\
\hline
2D model (homogeneous Owens  \cite{Owens2006})                                                                    & 1.0                                                          & 0.50                                                            & 0\%                                                               &  \\
Pott et al. \cite{Pott1997}                                                              & 0.8                                                          & 0.70                                                            & 29\%                                                              &  \\
Valdueza et al. \cite{Valdueza1997}                                                          & 1.7                                                          & 0.59                                                            & 15\%                                                              &  \\
\begin{tabular}[c]{@{}l@{}}Lindegaard et al. \cite{Lindegaard1987}, Patient 1\end{tabular} & 1.07                                                         & 0.55                                                            & 9\%                                                               &  \\
\begin{tabular}[c]{@{}l@{}}Lindegaard et al. \cite{Lindegaard1987}, Patient 5\end{tabular} & 1.13                                                         & 0.59                                                            & 15\%                                                              & \\
\hline
\end{tabular}
\end{table}
\section{N-Owens and I-Owens Schemes}\label{sec:8}
The main challenge of these numerical simulations is to develop a robust and efficient numerical scheme to obtain an accurate numerical solution at values of practical interest of the Weissenberg and Reynolds numbers. One of the greatest challenges for numerical simulations using the Owens model is the presence of spurious oscillations during the numerical simulation. The most likely origin of the oscillations is the choice of the numerical method, leading to spatial discretization errors, especially near the regions of high curvature in our test case geometry, where the neck of the aneurysm sac meets the upper channel wall. This gives rise to spurious negative eigenvalues, causing the conformation tensor to lose its symmetric positive definiteness (SPD) and Hadamard instabilities to grow.

\begin{figure}[h]
\centering
\includegraphics[scale=0.65]{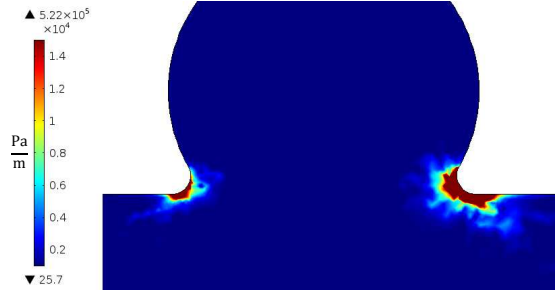}
\caption{The magnitude of $\nabla\cdot\tau_p$ in aneurysm region}
\label{Fig13}
\end{figure}

There have been several attempts by different researchers to conquer the high Weissenberg number problem, by transforming classical constitutive equations of Maxwell or Oldroyd type to new mathematical models to preserve the SPD of the conformation tensor. Examples include logarithm or hyperbolic tangent representations \cite{Fattal2004, Jafari2010,Jafari2018}. 
In our simulations, we observe that the value of the divergence of the polymeric contribution to the Cauchy stress tensor $\nabla\cdot\tau_P$ increases strongly close to the corners of the neck of the aneurysm sac where the curvature of the boundary is high, Fig. \ref{Fig13}.\\

\begin{figure}
\centering
\includegraphics[scale=0.75]{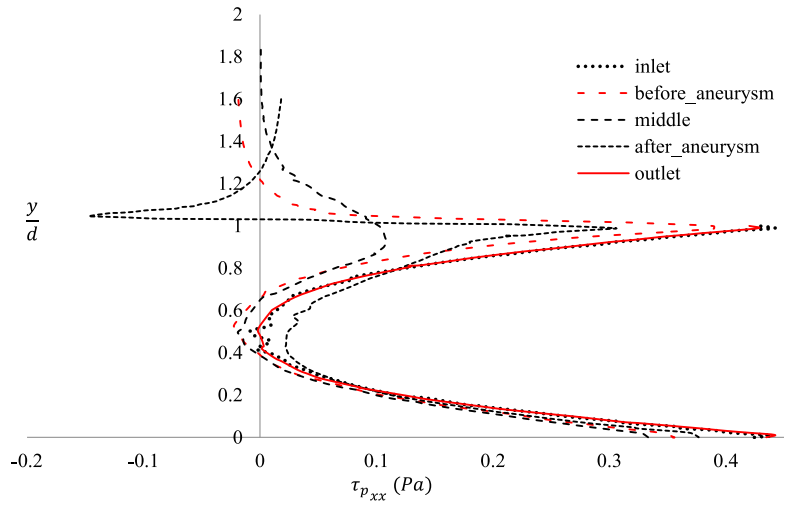}
\caption{${\tau_p}_{xx}$ in vertical cross section of the vessel at different positions}
\label{Fig14}
\end{figure}

\begin{figure}
\centering
\includegraphics[scale=0.75]{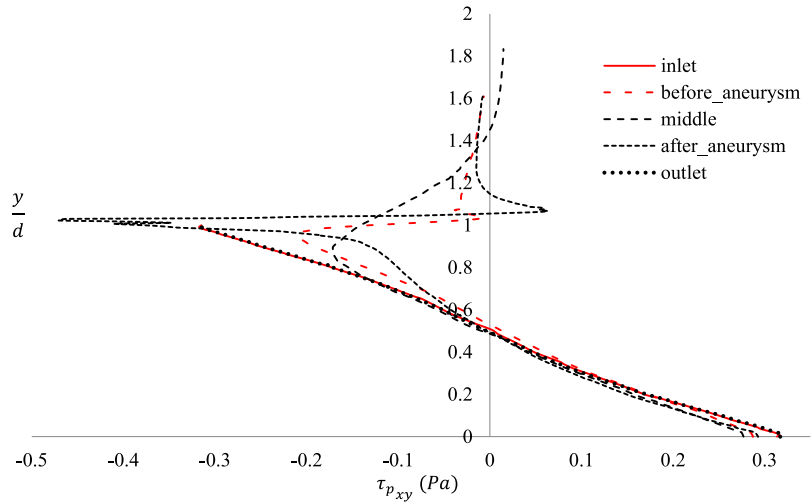}
\caption{${\tau_p}_{xy}$ in vertical cross section of the vessel at different positions}
\label{Fig15}
\end{figure}

\begin{figure}
\centering
\includegraphics[scale=0.75]{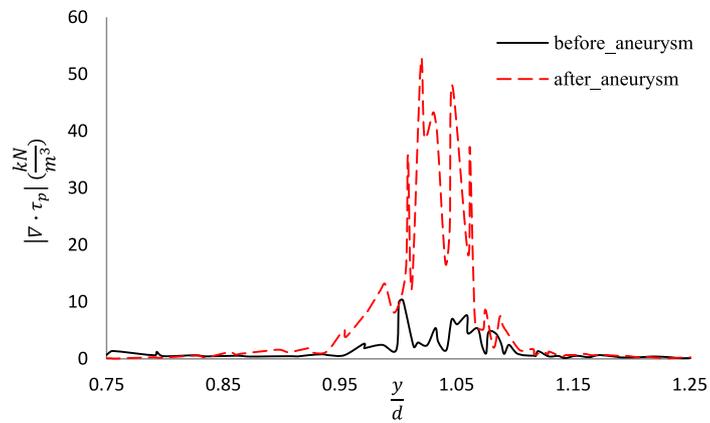}
\caption{The variation of $\nabla\cdot\tau_p$ in vertical cross section of the vessel before and after the aneurysm}
\label{Fig16}
\end{figure}

The profiles of ${\tau_p}_{xx}$ and ${\tau_p}_{xy}$ are shown in Fig. \ref{Fig14} and Fig. \ref{Fig15} before and after the aneurysm sac. Values of both ${\tau_p}_{xx}$ and ${\tau_p}_{xy}$ are changing rapidly in a small region around the aneurysm neck and this explains increasing the high magnitude of $\nabla\cdot\tau_p$  in that region.\\
Fig. \ref{Fig16} shows that the variation of $\nabla\cdot\tau_p$ changes dramatically before and after aneurysm near the aneurysm neck.
To simplify the problem, we replaced the contribution of the total stress (due to the viscous and polymeric stresses), $\nabla\cdot\tau$, in the momentum equation with a term of Newtonian type and adjusted the apparent viscosity, $\mu_e$ to include non-Newtonian effects (presence of the RBCs, adding to the total viscosity). With this assumption $\nabla\cdot\tau$ was replaced by $\mu_e\dot{\gamma}$ where $\mu_e$ was defined as the high shear-rate limit of
\begin{equation}
\mu _{e} = \frac{ \mu _{ \infty } }{ \lambda _{H} }  \Lambda + \mu _{s},
\label{eq09}
\end{equation}
$\mu_s$ being the plasma viscosity and $\Lambda$ a function collapsing to the relaxation time of a single red cell $\lambda_H$ as the shear rate tends to infinity. Thus, in the high shear rate regime $\mu_e$ was taken to be $\mu_{\infty}+\mu_s$ which is equal to $0.00334 [Pa.s]$ and close to the viscosity of Newtonian blood as mentioned in Eq. (\ref{eq03}).

\begin{figure}
\centering
\includegraphics[scale=0.75]{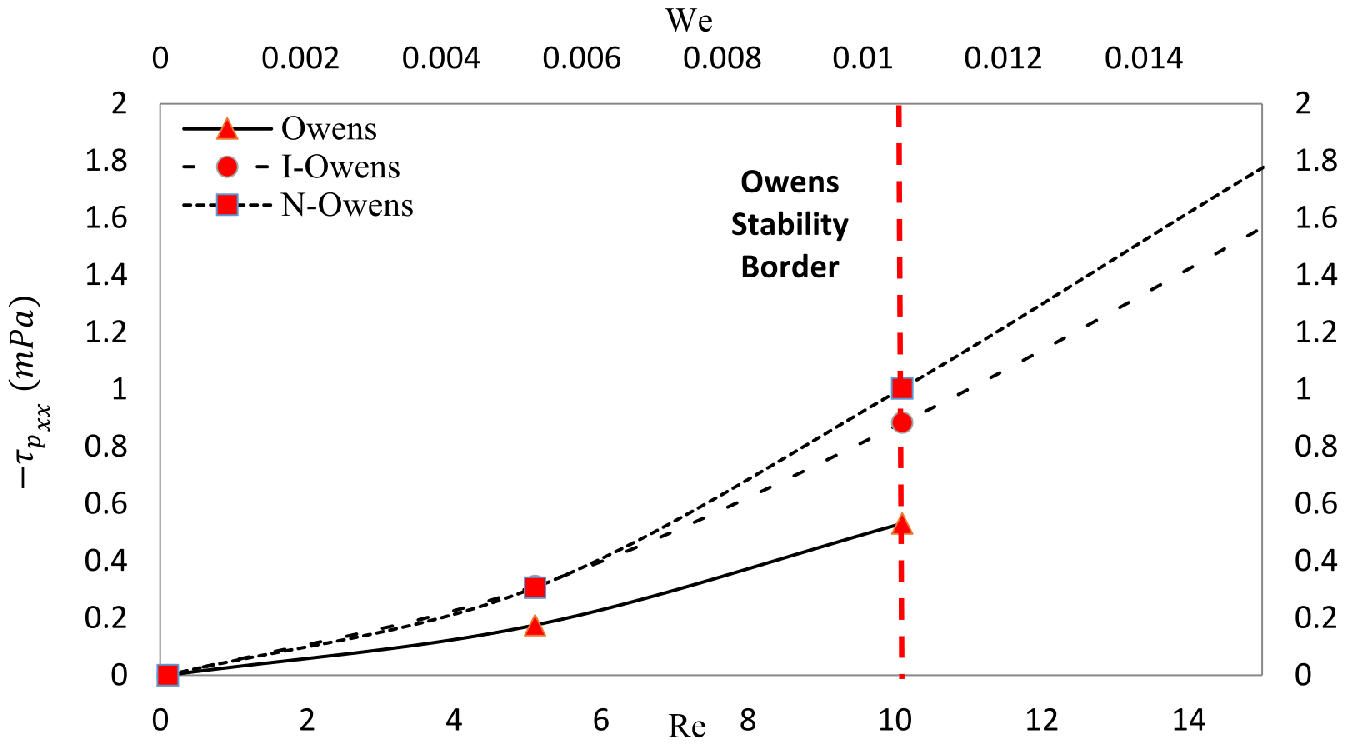}
\caption{The value of ${\tau_p}_{xx}$ at different values of $Re$ and $We$ with the Owens, I-Owens and N-Owens models in an aneurytic channel}
\label{Fig17}
\end{figure}

\begin{figure}
\centering
\includegraphics[scale=0.75]{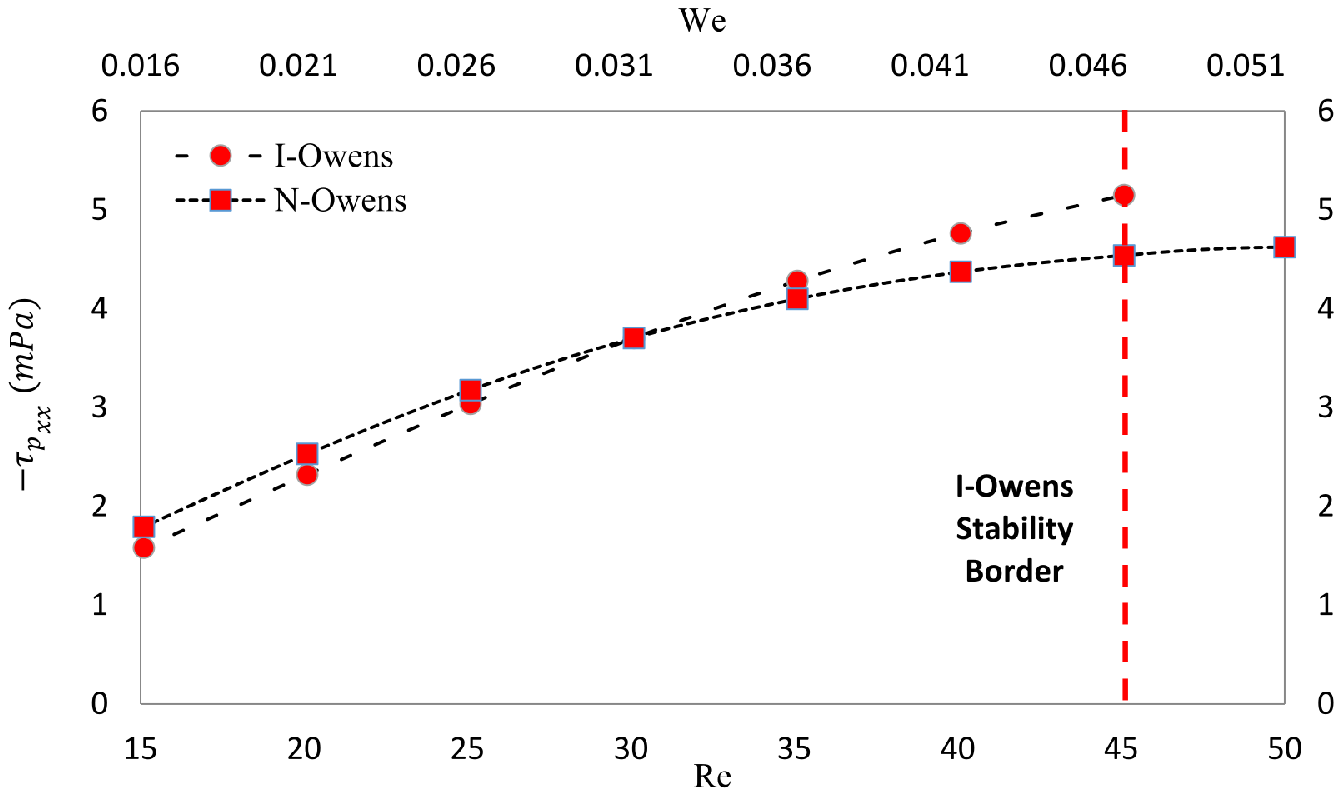}
\caption{The value of ${\tau_p}_{xx}$ at different values $Re$ and $We$ with the Owens, I-Owens and N-Owens models in an aneurytic channel}
\label{Fig18}
\end{figure}

\begin{figure}
\centering
\includegraphics[scale=0.75]{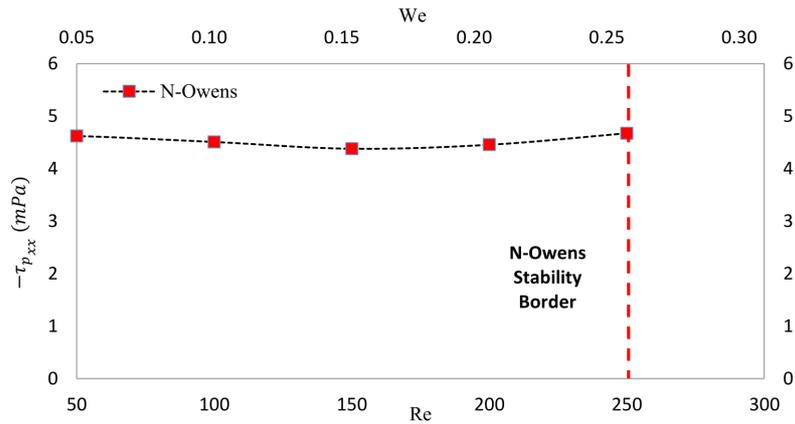}
\caption{The value of ${\tau_p}_{xx}$ at different values of $Re$ and $We$ with the Owens, I-Owens and N-Owens models in an aneurytic channel}
\label{Fig19}
\end{figure}

The value of $\nabla\cdot\tau_p$ increases with increasing Reynolds numbers so at low values of $Re$ the effect of polymeric stress on the blood velocity is small and its effect only becomes important at high Reynolds numbers.  At high values of $Re$ (or $We$) the linear momentum equations may be solved using a Newtonian stress term in order to determine the velocity field and the polymeric stresses then evaluated by using the constitutive equation of the Owens model. This procedure can improve the stability of the Owens model. We call this way of coupling the constitutive equation to the linear momentum equation the N-Owens scheme, with N indicating that we use the Newtonian velocity field calculated with a constant effective viscosity. 
The results of the N-Owens scheme for steady flow are presented in Figs.\ref{Fig17} to \ref{Fig19} for ${\tau_p}_{xx}$. The position where we measure the magnitude of ${\tau_p}_{xx}$ in our geometry is shown in Fig. \ref{Fig1}\\
As may be seen in Fig. \ref{Fig19}, the N-Owens scheme increases the critical Weissenberg number to about $17$ times that of the classical Owens model in the aneurytic channel. The predictions of the N-Owens scheme deviate from those of the original Owens model mostly at low  values of $Re$, as is to be expected. As mentioned above, the N-Owens scheme was designed for high shear-rate values where the instabilities grow, so that at low values of $|\dot{\gamma}|$,  the N-Owens scheme results deviate from those of the Owens model because the Newtonian equivalent viscosity is no longer the correct one valid for $|\dot\gamma| <<1$. We defined an alternative scheme called I-Owens to increase stability of the Owens model without modification to the original constitutive-momentum equations. ``I'' stands for initialization of the Owens model with the N-Owens model. The normal stress ${\tau_p}_{xx}$ in steady flow of the I-Owens scheme is shown in Figs.\ref{Fig17} and \ref{Fig18}. Initializing the numerical simulations with N-Owens can increase the attainable critical Weissenberg and Reynolds numbers compared to solving the constitutive and momentum equations with the initial and boundary conditions. \\
\section{Conclusions}
In the present work we have investigated the effect of different haemorheological models on the haemodynamics of blood flow through an aneurytic channel for both steady and pulsatile flows. As expected, increasing the Reynolds number decreases the impact of viscoelasticity on the blood flow. The Owens and Newtonian models approximately predict the same velocity profile in both steady and pulsatile flow with, however, significantly different $WSS$ near the neck of the aneurysm sac. This difference stems from the effect of the polymeric stress in the Owens constitutive equation. The velocity field of the C-Y model has significant differences from that of the Newtonian model at low Reynolds numbers but these differences become negligible at high Reynolds numbers. The C-Y model predicts $WSS$ values somewhere between the values computed using the Owens and Newtonian models. Pulsatility leads to smaller maximum values in velocity magnitude compared to those under steady state conditions.\\ 

Numerical instabilities limited our simulations with the Owens model to flows at modest values of the Reynolds and Weissenberg numbers. To stabilize the model, we have proposed two new procedures for coupling the constitutive and momentum equations, the so-called N-Owens and I-Owens schemes. The idea behind the N-Owens scheme is to replace $\nabla\cdot\tau_p$ with $\mu_e\Delta u$ where $\mu_e$ is the effective viscosity that includes the viscosity due to the red blood cells. N-Owens can reduce the numerical instabilities present due to the coupling of the momentum and Owens constitutive equations and increase the critical values of $Re$ and $We$ to values $15-25$ times greater than the original ones. Another method that we have introduced is I-Owens which uses the N-Owens results to create initial conditions for the original problem so that the instabilities can be attenuated.The maximum attainable Reynolds and Weissenberg numbers have been shown to be $1.25-3$ times larger than those attainable using the Owens model with the original initial and boundary values.

\clearpage
\section*{Appendix A}
The constitutive equation of the Owens model \cite{Owens2006} was given in Eq.\ref{eq06} as 
\begin{equation}
\tau +  \lambda  \left( \frac{ \partial  \tau }{ \partial t} +  v  .  \nabla  \tau -  \nabla  v  .  \tau -  \tau  .   \nabla  v ^{T}  \right) =  \frac{  \mu _{ \infty } }{  \lambda _{H} }  \lambda  \dot{ \gamma}.
\end{equation}

In the above, $\lambda$ is the fluid relaxation time, defined by 
\setcounter{equation}{0}
\renewcommand\theequation{A.\arabic{equation}}
 \setcounter{table}{0}
\renewcommand\thetable{A.\arabic{table}}
\begin{equation}
\lambda =  \frac{n  \lambda _{H} }{1+ \left(0.5b \left( |\dot{ \gamma}|\right)n \left(n-1\right)+a \left( |\dot{ \gamma}|  \right) \right)  \lambda _{H}},
\label{appendixA_01}
\end{equation}
where the aggregation rate $a(|\dot{ \gamma}|)$ is modelled by 
\begin{equation}
a \left( |\dot{ \gamma }| \right)=\left\{\begin{matrix} a_{1,3} |\dot{ \gamma }|^{3}+ a_{1,2} |\dot{ \gamma }|^{2}+ a_{1,0} \quad \hfill 0 \leq|\dot{ \gamma }| \leq  |\dot{ \gamma }_{c}|,\\ a_{2,3}|\dot{ \gamma }|^{3}+ a_{2,2} |\dot{ \gamma }|^{2}+ a_{2,1}|\dot{ \gamma }| + a_{2,0} \quad \hfill |\dot{ \gamma }|_{c} \leq |\dot{ \gamma }| \leq  |\dot{ \gamma }|_{max},\\ 0 \quad \hfill |\dot{ \gamma }|_{max} <  |\dot{ \gamma }|,  \end{matrix}\right.
\label{appendixA_02}
\end{equation}
for a certain choice of fitted parameters $a_{1,0}$, $a_{1,2}$, $a_{1,3}$, $a_{2,0}$, $a_{2,1}$, $a_{2,2}$, $a_{2,3}$. $|\dot{ \gamma }|_{c} $ and $|\dot{ \gamma }|_{max}$ denote, respectively, critical and maximum values of the shear rate affecting the aggregation rate. The disaggregation rate $b(|\dot{ \gamma}|)$ takes the form 

\begin{equation}
b \left( |\dot{ \gamma }| \right) =  \frac{a \left( |\dot{ \gamma } |\right)}{n_{st} \left(n_{st}-1 \right)  },
\label{appendixA_08}
\end{equation}
and appears in the differential equation satisfied by the average aggregate size $n$:
\begin{equation}
\frac{dn}{dt} = -\frac{1}{2}b \left( |\dot{ \gamma }| \right)(n-n_{st})(n+n_{st}-1),
\end{equation}
where 
\begin{equation}
n_{st}:=  \frac{ \mu _{0} }{\mu _{ \infty }}  \left(1+1.5a \left( |\dot{ \gamma }| \right) \lambda _{H}\right)  \left( \frac{1+ \phi | \dot{ \gamma}| ^{m}}{1+ \beta|\dot{ \gamma}| ^{m}} \right),
\label{appendixA_09}
\end{equation}
is the steady-state value of $n$. Finally, in (\ref{appendixA_09}), the parameters $\phi$ and $\beta$ are chosen so that 
\begin{equation}
\frac{ \phi }{ \beta } =  \frac{ \mu _{ \infty } }{\mu _{0}}.
\label{appendixA_10}
\end{equation}
Table \ref{TableA1} gives the parameter values used in our simulations. 
\begin{table}[h]
\centering
\caption {Parameters of the Owens model} \label{TableA1} 
\begin{tabular}{ | m{2cm} | m{2cm}| } 
\hline
Parameters & Values\\ 
\hline
$\mu_s$& $0.001 [Pa.s] $ \\ 
$\mu_0$ &  $0.0326 [Pa.s] $\\
$\mu_{\infty}$ &  $0.00234 [Pa.s] $\\ 
$\beta$ &  $0.7014 [s^m]$\\ 
$m$ &  $2.1238$\\ 
$\lambda_H$ &  $0.005 [s] $\\ 
$\dot\gamma_{max}$ &  $900 [s^{-1}]$\\ 
$\dot\gamma_c$ &  $5.78 [s^{-1}] $\\ 
\hline
\end{tabular}
\end{table}

\clearpage
\section*{Appendix B}
\setcounter{equation}{0}
\renewcommand\theequation{B.\arabic{equation}}
\setcounter{table}{0}
\renewcommand\thetable{B.\arabic{table}}
Here we explain the derivation of boundary conditions for the Owens model. To find the appropriate boundary conditions we assumed steady-state and fully-developed conditions, so that the $x-$momentum and polymeric stress components were expressed as follows:
\begin{equation}
\frac{ \partial p}{ \partial x} =  \frac{ \partial \tau _{xy} }{ \partial y} +  \mu _{s}  \frac{ \partial ^{2}v_{x}}{\partial y ^{2} },
\label{appendixB_01}
\end{equation}
\begin{equation}
\tau _{xy}=  \frac{ \lambda   \mu _{ \infty } }{  \lambda _{H} } \frac{\partial v_{x}}{\partial y},
\label{appendixB_02}
\end{equation}
\begin{equation}
\tau _{xx}=  2 \lambda \tau _{xy}\frac{\partial v_{x}}{\partial y},
\label{appendixB_03}
\end{equation}
\begin{equation}
\tau _{yy}=0.
\label{appendixB_04}
\end{equation}
In steady-state and with the assumption of large shear rates, the velocity field and extra stresses at inflow were expressed as in Eqs \ref{appendixB_05} to \ref{appendixB_06}
\begin{equation}
\ v _{x}=0,\; 
\ v _{y}= \frac{4Re  \mu _{N} }{ \rho d} \left(1- \frac{y}{d}\right)  \left( \frac{y}{d} \right),
\label{appendixB_05}
\end{equation}
\begin{equation}
\tau _{xy} = \frac{ \lambda   \mu _{ \infty } }{ \lambda _{H} }  \frac{4Re  \mu _{N} }{ \rho d^{2} } \left(1- \frac{2y}{d}\right),
\tau _{xx} = \frac{ 2  \lambda^{2} \mu _{ \infty } }{ \lambda _{H} } \left(\frac{4Re  \mu _{N} }{ \rho d^{2} } \left(1- \frac{2y}{d}\right)\right) ^2.
\label{appendixB_06}
\end{equation}

\clearpage 

\bibliographystyle{ieeetr}
\bibliography{paper1}

\end{document}